\documentclass[12pt]{iopart}

\pdfoutput=1

\usepackage{graphicx}
\usepackage{xfrac}
\usepackage{epsfig}
\usepackage{iopams}
\usepackage{color}

\begin{document}

\title{Bose-Hubbard dynamics of polaritons in a chain of circuit QED cavities}

\author{Martin Leib and Michael J. Hartmann}
\address{Technische Universit{\"a}t M{\"u}nchen, Physik Department, James-Franck-Str., 85748 Garching, Germany}
\eads{\mailto{martin.leib@ph.tum.de},\mailto{michael.hartmann@ph.tum.de}}

\date{\today}

\begin{abstract}
We investigate a chain of superconducting stripline resonators, each interacting with a transmon qubit, that are capacitively coupled in a row. We show that the dynamics of this system can be described by  a Bose-Hubbard Hamiltonian with attractive interactions for polaritons, superpositions of photons and qubit excitations. This setup we envisage constitutes one of the first platforms where all technological components that are needed to experimentally study chains of strongly interacting polaritons have already been realized. By driving the first stripline resonator with a microwave source and detecting the output field of the last stripline resonator one can spectroscopically probe properties of the system in the driven dissipative regime. We calculate the stationary polariton density and density-density correlations $g^{(2)}$ for the last cavity which can be measured via the output field. Our results display a transition from a coherent to a quantum field as the ratio of on site interactions to driving strength is increased.
\end{abstract}


\submitto{\NJP}
\maketitle 

\tableofcontents

%
%
\section{Introduction}
In recent years, the investigation of condensed matter and quantum many-body systems with quantum simulators, artificial quantum many-body systems that offer unprecedented controllability and measurement access in the laboratory, has become an active research area and is currently receiving increasing attention. The technology employed for quantum simulators ranges from ultra-cold atoms \cite{review_Bloch_2008} to ion traps \cite{Friedenauer:2008fk} and systems of coupled cavities \cite{Hartmann:2006kx,hartmann-2008-2,PhysRevA.76.031805,GTH06} among others. 

Due to recent technological progress, arrays of coupled cavities and optical nano-fibers in which the trapped light modes couple to atoms are now becoming suitable devices for the generation of quantum many-body systems of polaritons \cite{Hartmann:2006kx,hartmann-2008-2,PhysRevA.76.031805,GTH06,PhysRevLett.99.103601,HBP08,MKiffner_1,MKiffner_2}, i.e. quantum-mechanical superpositions of atomic and photonic excitations. In these systems, it is of particular interest but also most challenging to reach a strongly correlated regime, where their dynamics differs most significantly from that of classical light fields.
The key experimental requirement for reaching these conditions is a so called strong coupling regime for the cavities respectively the fiber.
This means that the coherent coupling between the light modes and the atoms or other optical emitters must be strong compared to the loss processes which are inevitably present in every device. A very impressive strong coupling regime has recently been realized in circuit cavities, making these devices an ideal platform for studying strongly correlated polaritons.

Circuit QED \cite{PhysRevA.69.062320,Wallraff:2004rz,DMM+08} was developed as a solid state equivalent to optical cavity QED, coupling Josephson qubits that are acting as artificial atoms with stripline resonators acting as cavities for microwave photons. Here, the reduced quasi one-dimensional mode volume of the stripline resonator and the enhanced dipole moment of the Josephson qubit with respect to atoms give rise to a pronounced strong coupling regime, where the coupling between resonator and qubit, $g$, significantly exceeds both the decay rate of the resonator, $\kappa$, and the qubit, $\gamma$, $g/\kappa\gg 1$ and $g/\gamma\gg 1$. Moreover, since stripline resonators trap microwave photons, they are more than 1cm long. The precision of current fabrication techniques thus allows to build several resonators that can resonantly couple to each other via mutual photon tunneling on the same chip \cite{2010arXiv1003.2734J,Schmidt2010,Koch2010}. 
The currently employed circuit QED technology thus permits to build arrays of resonantly coupled cavities that each interact in a strong coupling regime with qubits. In this way it is one of the first setups to feature all properties which are needed to generate strongly correlated many-body systems of polaritons.

In this work we show that an effective Bose-Hubbard Hamiltonian for polaritons can be engineered in an array of stripline resonators that each couple to a transmon qubit \cite{koch:042319}.
Josephson qubits \cite{RevModPhys.73.357,devoret-2004} come in basically three different flavours depending on the property that is controlled from the outside or rather the channel of the qubit environment coupling: flux-\cite{flux:Qubit}, phase-\cite{PhysRevLett.89.117901} and charge- qubits \cite{PhysRevB.36.3548,Nakamura:1999uq}. Here we consider a setup with transmon qubits \cite{koch:042319} which are charge qubits (Cooper pair boxes) operated at sufficiently enhanced values for the ratio of Josephson energy, $E_{J}$, over charging energy, $E_{C}$, $E_J/E_C\geq50$ and are robust against decoherence caused by fluctuations of background charges. We emphasize that the effective Bose-Hubbard Hamiltonian we derive can be realized with resonators and qubits of readily existing technology. 

Quantum phases for the ground state and low temperature thermal states of the Bose-Hubbard Hamiltonian have been studied with ultra-cold atoms trapped in optical lattices \cite{review_Bloch_2008}. This system has also been employed to study the dynamics of none-equilibrium states that were prepared by sudden quenches of some lattice parameters \cite{2010Sci...327.1621H}. In contrast, a realisation in an array of stripline resonators allows to investigate the Bose-Hubbard Hamiltonian in a fundamentally different regime, where the resonator array is permanently driven by lasers to load it with photons and thus compensate for the excitations that are lost due to qubit relaxation and cavity decay. Whereas substantial understanding of equilibrium quantum phase transitions has been achieved, a lot less is known about these non-equilibrium scenarios where the dynamical balance between loading and loss mechanisms leads to stationary states. It is the investigation of these stationary states, that our approach to the Bose-Hubbard Hamiltonian is ideally suited for.
 
Experiments with transmon qubits \cite{koch:042319,Majer:2007sf} coupled to a stripline resonator are often conducted without directly measuring the state of the qubit but by spectroscopically probing the transmission properties of the resonator. 
In an experiment the effective Bose-Hubbard Hamiltonian will thus be operated out of thermal equilibrium in a driven dissipative regime \cite{PhysRevLett.104.113601,Gerace:2009kx,PhysRevLett.103.033601,2009arXiv0904.4437T}. 
In a suitable setup with a linear chain of resonators one would thus drive the first resonator with a coherent microwave input and measure the properties of the output signal at the opposite end of the chain.

In the regime we consider, this situation can be accurately described by a Bose-Hubbard Hamiltonian for polaritons with a coherent driving term at the first site and Markovian losses of polaritons due to cavity decay and qubit relaxation at all sites of the chain. In this scenario, the interplay of coherent drive and polariton loss leads to the emergence of steady states, for which we derive the particle statistics and characteristic correlations. In doing so we focus on the polariton statistics, in particular the density and density-density correlations, in the last resonator as these can be measured via the output signal. Our results show a transition from a coherent field to a field with strongly non-classical particle statistics as the ratio of on site interactions to driving strength is increased.

This work is devided into two main parts. In \sref{sec:BH} we show that the dynamics of our system can be described by a Bose-Hubbard model and in \sref{sec:g2} we present the results of our calculations for the polariton density and density-density correlations.

\section{Transmon-QED and the Bose-Hubbard Model}\label{sec:BH}

\begin{figure}
\includegraphics{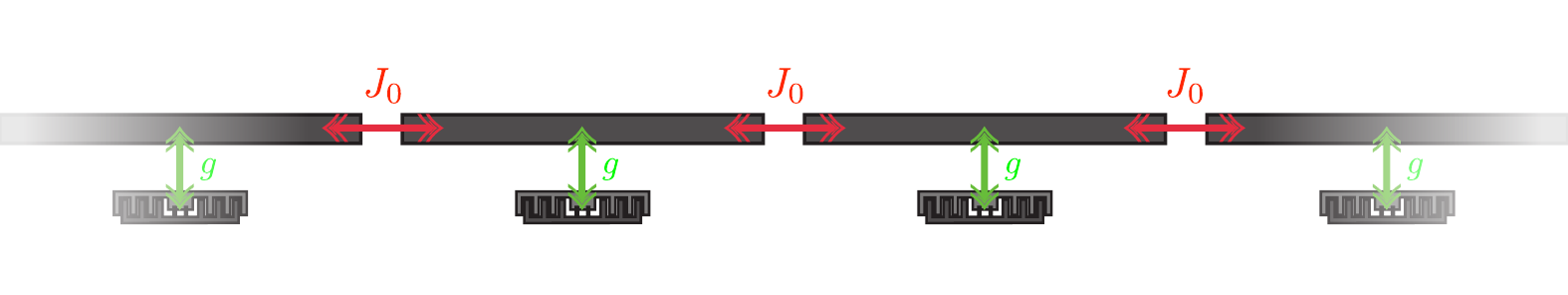}
\caption{Sketch of the proposed system to simulate Bose-Hubbard physics. Stripline resonators are coupled capacitively in a chain and each stripline resonator is coupled to a transmon qubit that gives rise to an on-site interaction for the polaritons. For definitions of $J_0$ and $g$, see \eref{eq:Coupling} and \eref{eq:Hamiltonian1}.}
\label{fig:fullHamiltonian}
\end{figure}

To generate a Bose-Hubbard model with polaritons we consider an array of capacitively coupled stripline resonators with each resonator coupled to a transmon qubit, see \fref{fig:fullHamiltonian}. In this section, we first introduce the Hamiltonian that describes this setup and then show how it can be considerably simplified and transformed into a Bose-Hubbard Hamiltonian for two polariton species.

\subsection{The full Hamiltonian}
The full Hamiltonian of our setup is a sum of single-site Hamiltonians, $H_{\mathrm{1-site},i}$, that each describe a transmon qubit coupled to a stripline resonator and terms that describe the capacitive coupling between neighbouring stripline resonators, $H_{C_J,i,i+1}$,
\begin{equation}\label{eq:fullHamiltonian}
H=\sum_i\left(H_{\mathrm{1-site},i}+H_{C_J,i,i+1}\right)\,.
\end{equation}
A transmon qubit can be regarded as a Cooper pair box that is operated at a large ratio of $E_J/E_C\gg 1$.
This regime can be accessed by shunting the Josephson junction with an additional large capacitance and thereby lowering the charging energy $E_C=e^2/(2C_{\Sigma})$. Here, $C_{\Sigma}= C_{J} + C_{g} + C_{B}$ is the sum of the junction's capacitance, $C_J$, the mutual capacitance with stripline resonator, $C_g$, and the shunting capacitance $C_B$.
The Hamiltonian for one stripline resonator coupled to a transmon qubit reads,
\begin{equation}\label{eq:1-siteHamiltonian}
H_{\mathrm{1-site}}=4E_C\left(\hat{n}-n_g^{dc}-n_g^{ac}\right)^2-E_J\cos\left(\hat{\varphi}\right)+\omega_r''a^{\dagger}a\,.
\end{equation}
Here, $\omega_r''$ is the resonance frequency of the isolated resonator and we have omitted the site-index $i$ for readability. Transmon qubits consist of two superconducting islands connected by Josephson junctions. $\hat{n}$ is the operator for the difference in the number of Cooper pairs on the two superconducting islands, $n_g^{dc}$ the offset charge induced by an applied dc voltage and intrinsical defects and $\hat{\varphi}$ is the operator for the superconducting-phase difference between the two islands. We assume the transmon qubit to be placed in the antinode of the stripline resonator's field mode. This gives rise to an additional ac component in the offset charge, $n_g^{ac}=\sfrac{C_g}{(2e)}V_{\mathrm{rms}}^{(0)}(a+a^{\dagger})$, where $V_{\mathrm{rms}}^{(0)}=\sqrt{\omega_r''/C_r}$ is the root mean square voltage of the vacuum field mode and $a$ the annihilation operator of photons in the stripline resonator. The offset charge $n_g^{ac}$ thus induces a coupling between the transmon and photons in the resonator. For circuit QED setups one normally uses $\lambda$-resonators with the antinode located at the middle of the resonator.

The energy of the coupling capacitor between neighbouring transmission line resonators, e.g. sites $i$ and $i+1$, can be expressed in terms of the difference in the electrostatic potentials across the capacitor,
\begin{equation*}
H_{C_J,i,i+1} = \frac{C_J(\hat{V}_i-\hat{V}_{i+1})^2}{2}=\frac{C_J}{C_r}\frac{\omega_r''}{2}(a^{\dagger}_i+a_i-a_{i+1}^{\dagger}-a_{i+1})^2\,.
\end{equation*}
Here, $C_r$ is the capacitance of the whole stripline resonator with respect to the ground plane and $C_J$ the capacitance of the capacitor that connects the two resonators. We assume the electrostatic potential in resonator $i$ to have antinodes at the ends of the resonator and write it in terms of the creation and annihilation operators, $a_{i}^{\dagger}$ and $a_{i}$.
We now turn to simplify the Hamiltonian \eref{eq:fullHamiltonian} by a sequence of approximations.

\subsection{Approximations to single-site terms}
We first simplify the single-site terms, $H_{\mathrm{1-site}}$, as in \eref{eq:1-siteHamiltonian}. For large $E_J/E_C$ and low energies, the phase difference between the two islands remains small and we can expand the cosine in \eref{eq:1-siteHamiltonian} around $\varphi=0$ up to quartic order, 
\begin{equation}\label{eq:TransmonQED}
H^{(1)}_{\mathrm{1-site}}=4E_C\left(\hat{n}-n_g^{dc}-n_g^{ac}\right)^2-E_J+\frac{E_J}{2}\hat{\varphi}^2-\frac{E_J}{24}\hat{\varphi}^4 +\omega_r'' a^{\dagger}a\,.
\end{equation}
Higher order terms can be neglected, c.f. \cite{koch:042319}.
In terms of bosonic creation and annihilation operators for the transmon qubit excitations,
\begin{eqnarray}
\hat{n}&=&\frac{i}{2}\left(\frac{E_J}{2E_C}\right)^{\frac{1}{4}}\left(b-b^{\dagger}\right)\nonumber\\
\hat{\varphi}&=&\left(\frac{2E_C}{E_J}\right)^{\frac{1}{4}}\left(b+b^{\dagger}\right)\nonumber
\end{eqnarray}
the Hamiltonian \eref{eq:TransmonQED} reads,
 \begin{equation}\label{eq:Hamiltonian1}
H_{\mathrm{1-site}}^{(1)}=H_{\mathrm{transmon}}+H_{\mathrm{lin}}+H_{\mathrm{coupling}}+H_{res}\,,\\
\end{equation}
where
\begin{eqnarray}
H_{\mathrm{transmon}}&=&\omega_qb^{\dagger}b-\frac{E_C}{12}(b+b^{\dagger})^4\nonumber\\
H_{\mathrm{lin}}&=&-i4E_C n_g^{dc}\left(\frac{E_J}{2E_C}\right)^{\frac{1}{4}}(b-b^{\dagger})+\nonumber\\
&&+2\frac{C_g}{C_{\Sigma}} e V_{rms}^{(0)}n_g^{dc}\left(a+a^{\dagger}\right)\nonumber\\
H_{\mathrm{coupling}}&=&ig(b-b^{\dagger})(a+a^{\dagger})\nonumber\\
H_{\mathrm{res}}&=&\omega_r' a^{\dagger}a.\nonumber
\end{eqnarray}
with,
\begin{eqnarray}
\omega_q&=&\sqrt{8E_CE_J}\nonumber\\
\omega_r'&=&\omega_r''\left(1+\frac{C_g^2}{C_{\Sigma}C_r}\right)\nonumber\\
g&=&\frac{C_g}{C_{\Sigma}} e V_{rms}^{(0)}\left(\frac{E_J}{2E_C}\right)^{\frac{1}{4}}\,.\nonumber
\end{eqnarray}
The terms linear in the creation and annihilation operators can be eliminated by performing the unitary transformation
\begin{eqnarray}
U&=&U_1\otimes U_2\nonumber\\
U_1&=&\exp\left(ra-ra^{\dagger}\right)\nonumber\\
U_2&=&\exp\left(i(sb^{\dagger}+sb)\right)\,.\nonumber
\end{eqnarray}
that displaces the creation and annihilation operators by the constants $r$ and $s$ respectively
\begin{eqnarray}
a&\to&U_1aU_1^{\dagger}=a+r\nonumber\\
b&\to&U_2bU_2^{\dagger}=b-is.\nonumber
\end{eqnarray}
$r$ and $s$ can now be chosen such that all terms linear in $a$ and $b$ cancel in the transformed Hamiltonian. Finally the interaction between the transmon qubit and the field mode of the stripline resonator gets reduced to an exchange interaction in a rotating wave approximation. To justify this rotating wave approximation we have to ensure that the interaction strength between the transmon qubit and the stripline resonator is smaller than the sum of the frequencies of the two,  
\begin{equation} \label{eq:RwaApprox}
\frac{g}{\omega_r'+\omega_q}\ll1\,.
\end{equation}
Parameters extracted from \cite{Fink:2008sf} are $\omega_r'=43.6 \mathrm{Ghz}$, $E_C=0.4 \mathrm{Ghz}$ and a maximal value for $E_J/E_C$ of $150$. We choose $C_g e V_{\mathrm{rms}}^{(0)}/(C_{\Sigma} \omega_r)=0.1$ which is in agreement with the theoretical upper bound in \cite{koch:042319} and find $\frac{g}{\omega_r'+\omega_q}\approx 0.1$. The single-site Hamiltonian can thus be approximated by,
\begin{equation}\label{eq:TransmonQED2}
H_{\mathrm{1-site}}^{(2)}=\sqrt{8E_CE_J}b^{\dagger}b-\frac{E_C}{12}\left(b+b^{\dagger}\right)^4
+g\left(a^{\dagger}b+ab^{\dagger}\right)+\omega_r'a^{\dagger}a\,.
\end{equation}

\subsection{Approximations to couplings between resonators}
We now turn to simplify the couplings between neighbouring resonators, $H_{C_J,i,i+1}$.
We assume that $C_J \ll C_r$ which implies that $C_J\omega_r''/(2C_r)$ is small compared to the isolated cavity frequency $\omega_r''$, $C_J/(2C_r) \ll 1$ and apply a rotating wave approximation to neglect those terms in the intercavity interaction that don't conserve the total photon number,
\begin{equation}\label{eq:Coupling}
H_{C_J,i,i+1}^{(1)} = J_0(a^{\dagger}_i a_i + a_{i+1} a^{\dagger}_{i+1})-J_0(a^{\dagger}_i a_{i+1} + a_i a^{\dagger}_{i+1})\,,
\end{equation}
where $J_0=\frac{C_J}{C_r}\omega_r'$. The first term on the rhs of \eref{eq:Coupling} can be absorbed into the single-site Hamiltonians by introducing a shifted resonator frequency
\begin{equation}\label{eq:resfrequ}
\omega_{r} = \omega_r'' \left(1 + \frac{C_g^2}{C_{\Sigma}C_r}+2\frac{C_J}{C_r}\left(1 + \frac{C_g^2}{C_{\Sigma}C_r}\right)\right)\,,
\end{equation}
and the remaining term in \eref{eq:Coupling} describes tunneling of photons between neighbouring resonators. Next, we explain how 
the simplified Hamiltonian $H^{(2)}=\sum_i\left(H^{(2)}_{1-site,i}+H_{C_J,i,i+1}^{(1)}\right)$ can be transformed to a two component Bose-Hubbard Hamiltonian.

\subsection{The polariton modes}
In the case of circuit QED with transmon qubits the coupling constant between photonic and qubit excitations is the dominating interaction energy of the system. Excitations of the whole system therefore can't be characterized as purely photonic or qubit excitations in general. To obtain a more suitable description we introduce new creation and annihilation operators,
\numparts
\begin{eqnarray}\label{eq:PolaritonModes}
c_+&=&\cos(\theta)a+\sin(\theta)b\\
c_-&=&\sin(\theta)a-\cos(\theta)b\,,
\end{eqnarray}
\endnumparts
describing excitations commonly termed polaritons where,
\begin{eqnarray}
\sin(\theta)&=&\frac{g}{\sqrt{g^2+\left(\Delta\omega+\sqrt{\Delta\omega^2+g^2}\right)^2}}\nonumber\\
\cos(\theta)&=&\frac{\Delta\omega+\sqrt{\Delta\omega^2+g^2}}{\sqrt{g^2+\left(\Delta\omega+\sqrt{\Delta\omega^2+g^2}\right)^2}}\,,\nonumber
\end{eqnarray}
with $\Delta\omega=\omega_r-\omega_q$.
The sine and cosine terms account for the transition of the character of the exitations from photonic to qubit excitations for the $c_+$-mode as the ratio $E_J/E_C$ increases and vice versa for the $c_-$-mode. 
 Expressing the Hamiltonian $H^{(2)}=\sum_i\left(H^{(2)}_{1-site,i}+H_{C_J,i,i+1}^{(1)}\right)$ in the polariton modes (\ref{eq:PolaritonModes}{\it-b}) we get,
\begin{equation*}
H^{(2)}= H_{c_+,\mathrm{lin}}+H_{c_-,\mathrm{lin}}+H_{\mathrm{cc}}+H_{\mathrm{nlin}}\,.
\end{equation*}
This Hamiltonian consists of two harmonic chains for the $c_+$ and $c_-$ polariton modes,
\begin{eqnarray}
H_{c_+,\mathrm{lin}}&=&\sum_i\left(\omega'_+c_{i,+}^{\dagger}c_{i,+}-J_0\cos^2(\theta)\left(c_{i,+}^{\dagger}c_{i+1,+}+h.c.\right)\right)\nonumber\\
H_{c_-,\mathrm{lin}}&=&\sum_i\left(\omega'_-c_{i,-}^{\dagger}c_{i,-}-J_0\sin^2(\theta)\left(c_{i,-}^{\dagger}c_{i+1,-}+h.c.\right)\right)\,,\nonumber
\end{eqnarray}
with $\omega'_{\pm}=((\omega_r+\omega_q)\pm\sqrt{\Delta^2+g^2})/2$, a term describing hopping from a $c_-$-mode at site $i$ to a $c_+$-mode at site $i+1$ and all other possible combinations,
\begin{equation*}
H_{\mathrm{cc}}=-J_0\sin(\theta)\cos(\theta)\sum_i\left(c_{i,+}^{\dag}c_{i+1,-}+c_{i,-}^{\dag}c_{i+1,+}+h.c.\right)\,,
\end{equation*}
and a term describing the nonlinearity,
\begin{equation*}
H_{\mathrm{nlin}}=\frac{-E_c}{12}\sum_i\left(\sin(\theta)\left(c_{i,+}+c_{i,+}^{\dagger}\right)-\cos(\theta)\left(c_{i,-}+c_{i,-}^{\dagger}\right)\right)^4\,.
\end{equation*}
We assume the frequencies of the two polariton modes to be well separated, apply another rotating wave approximation where we neglect the term $H_{\mathrm{cc}}$ and convert the nonlinearity term $H_{nlin}$ into Kerr form and get a renormalization of the polariton frequency and a density-density coupling between the polariton modes.
This requires the difference in frequencies for the unperturbed modes $\omega_+'-\omega_-' = \sqrt{\Delta\omega^2+g^2}$ involved to exceed the magnitude of the coupling between the modes and the nonlinearity,
\begin{eqnarray*}
\sqrt{\Delta\omega^2+g^2}&\gg& J_0\\
\sqrt{\Delta\omega^2+g^2}&\gg& \frac{E_C}{12}\,.
\end{eqnarray*}
Plugging in realistic values for the parameters, extracted for example from \cite{Fink:2008sf}, ($E_C=0.4 \mathrm{Ghz}$, $\omega_r=43,6 \mathrm{Ghz}$ $E_J/E_C=0..150$) we realize that the second inequality is indeed fulfilled. Engineering the capacitance $C_J$ such that $J_0$ is of the order of $E_C/12$, we can ensure that the first equality is fulfilled as well. The rotating wave approximation eliminates the intermode exchange coupling and we obtain a Bose-Hubbard Hamiltonian for both modes, $c_+$ and $c_-$, with a density-density coupling between them,
\begin{equation} \label{eq:BHHamiltonian}
H^{(3)}=H_{c_+}+H_{c_-}+H_{dd} \,,\\
\end{equation}
where
\begin{eqnarray}
H_{c_+}&=&\sum_i\Big(\omega_+c_{i,+}^{\dagger}c_{i,+}-J_+\left(c_{i,+}^{\dagger}c_{i+1,+}+h.c.\right)-\nonumber\\
&&-\frac{U_+}{2}c_{i,+}^{\dagger}c_{i,+}^{\dagger}c_{i,+}c_{i,+}\Big)\nonumber\\
H_{c_-}&=&\sum_i\Big(\omega_-c_{i,-}^{\dagger}c_{i,-}-J_-\left(c_{i,-}^{\dagger}c_{i+1,-}+h.c.\right)-\nonumber\\
&&-\frac{U_-}{2}c_{i,-}^{\dagger}c_{i,-}^{\dagger}c_{i,-}c_{i,-}\Big)-\nonumber\\
H_{\mathrm{dd}}&=&-\sum_i 2U_{+-}c_{i,+}^{\dagger}c_{i,+}c_{i,-}^{\dagger}c_{i,-} \,,\nonumber
\end{eqnarray}
with
\begin{eqnarray}
\omega_+&=&\omega'_++E_C\left(\cos^4(\theta)+\sin^2(\theta)\cos^2(\theta)\right)\nonumber\\
\omega_-&=&\omega'_-+E_C\left(\sin^4(\theta)+\sin^2(\theta)\cos^2(\theta)\right)\nonumber\\
J_+&=&J_0\cos^2(\theta)\nonumber\\
U_+&=&E_C\sin^4(\theta)\nonumber\\
J_-&=&J_0\sin^2(\theta)\nonumber\\
U_-&=&E_C\cos^4(\theta)\nonumber\\
U_{+-}&=&E_C\sin^2(\theta)\cos^2(\theta)\,.\nonumber
\end{eqnarray}
We thus arrived at a two component Bose-Hubbard model for the modes $c_+$ and $c_-$ with attractive interactions and a density-density coupling between both species. The two species are a mixture of stripline resonator field mode and qubit excitations (\ref{eq:PolaritonModes}{\it-b}) with different weights of the photonic or qubit contribution depending on the value of $E_J/E_C$. 

For small values of $E_J/E_C$ the $c_+$ polaritons become increasingly photonic. Consequently, their tunneling rate $J_{+}$ approaches the tunneling rate of bare photons, $J_{0}$, and their on-site interaction $U_{+}$ vanishes. For large $E_J/E_C$, on the other hand, $J_{+}$ vanishes and the nonlinearity $U_{+}$ approaches the nonlinearity of the qubits, $E_{C}$.
For the $c_-$ polaritons, the roles of both limits are interchanged.

\begin{figure}
\includegraphics{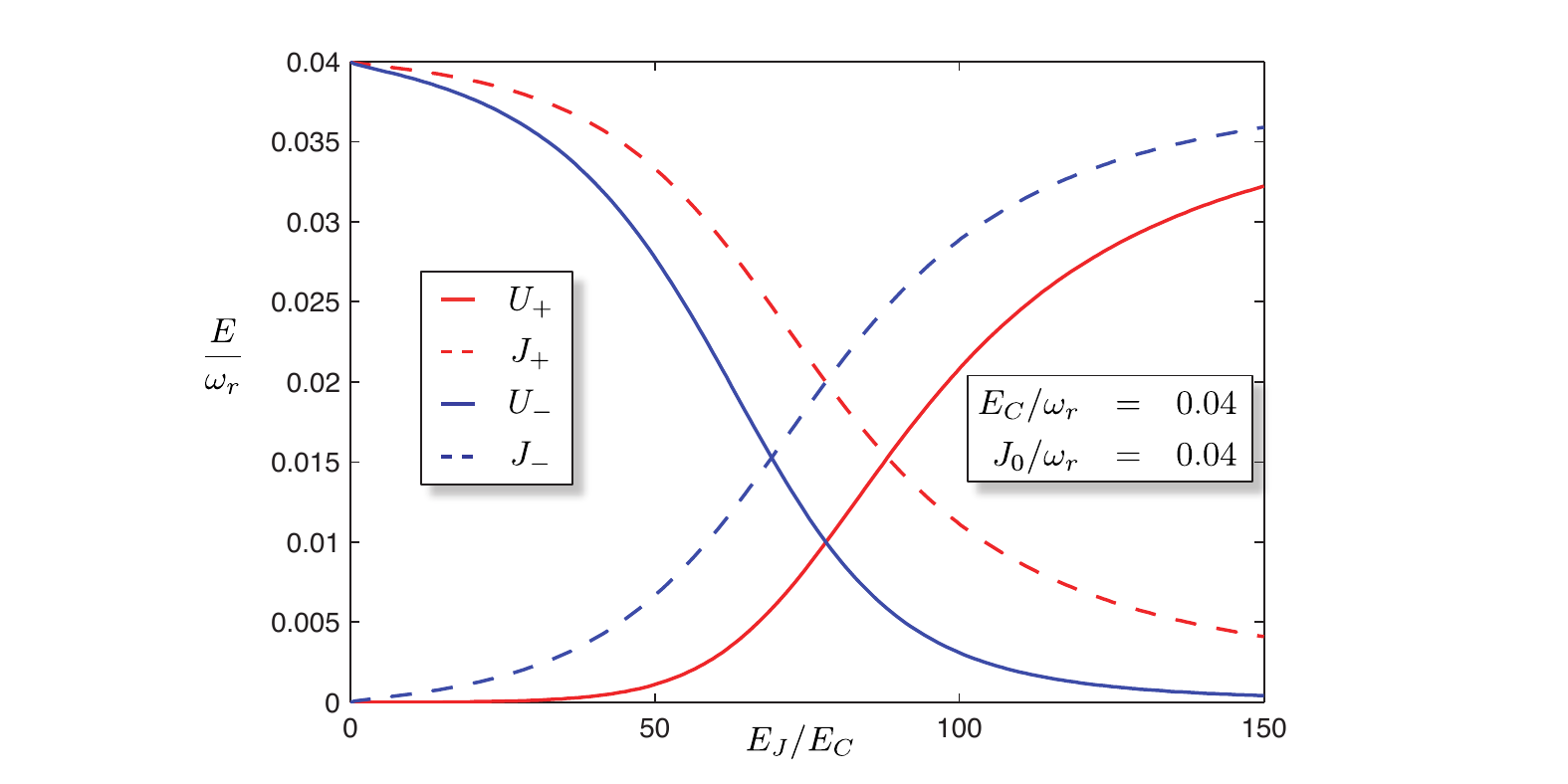}
\caption{Plot of the effective Bose-Hubbard parameters for the $c_+$ polariton mode, $U_+$ and $J_+$, and the $c_-$ polariton mode, $U_-$ and $J_-$. As the character of the polariton modes change as a function of $E_J/E_C$ from resonator field mode to qubit excitation for the $c_+$ polariton modes and vice versa for the $c_-$ polariton modes the Bose-Hubbard parameters change accordingly.}
\label{fig:BHparameters}
\end{figure} 
For each value of $E_J/E_C$, the separation between the resonance frequencies of $c_+$ and $c_-$ polaritons, $|\omega_{+} - \omega_{-}|$, is sufficiently large such that, in a scenario where we drive the first cavity by a microwave source, we can always adjust the frequency of the drive to only selectively excite one of the modes. For reasons that will become clear later we therefore choose the $c_+$-polaritons to be our quantum simulator for a driven dissipative Bose-Hubbard model.

\subsection{Validity of the approximations}
To further illustrate the validity of our approximations we compare the eigenenergies of the full Hamiltonian $H$, c.f. \eref{eq:fullHamiltonian}, approximated under assumption \eref{eq:RwaApprox}, with the eigenenergies of the Bose-Hubbard Hamiltonian $H^{(3)}$, c.f. \eref{eq:BHHamiltonian}. The single-site Hamiltonians summed up in the full Hamiltonian describe the interaction between transmon qubit and stripline resonator in a rotating wave approximation. This Hamiltonian has already been used to describe an experiment revealing the nonlinear response of a resonator and transmon qubit system with excellent agreement between theory and experimental data \cite{Bishop:2009kx}. Therefore comparison of the eigenvalues of our Bose-Hubbard Hamiltonian and the eigenvalues of the full Hamiltonian provides a good means to estimate the effects of the approximations we made. For simplicity we restricted our model to two sites.

\begin{figure}
\includegraphics{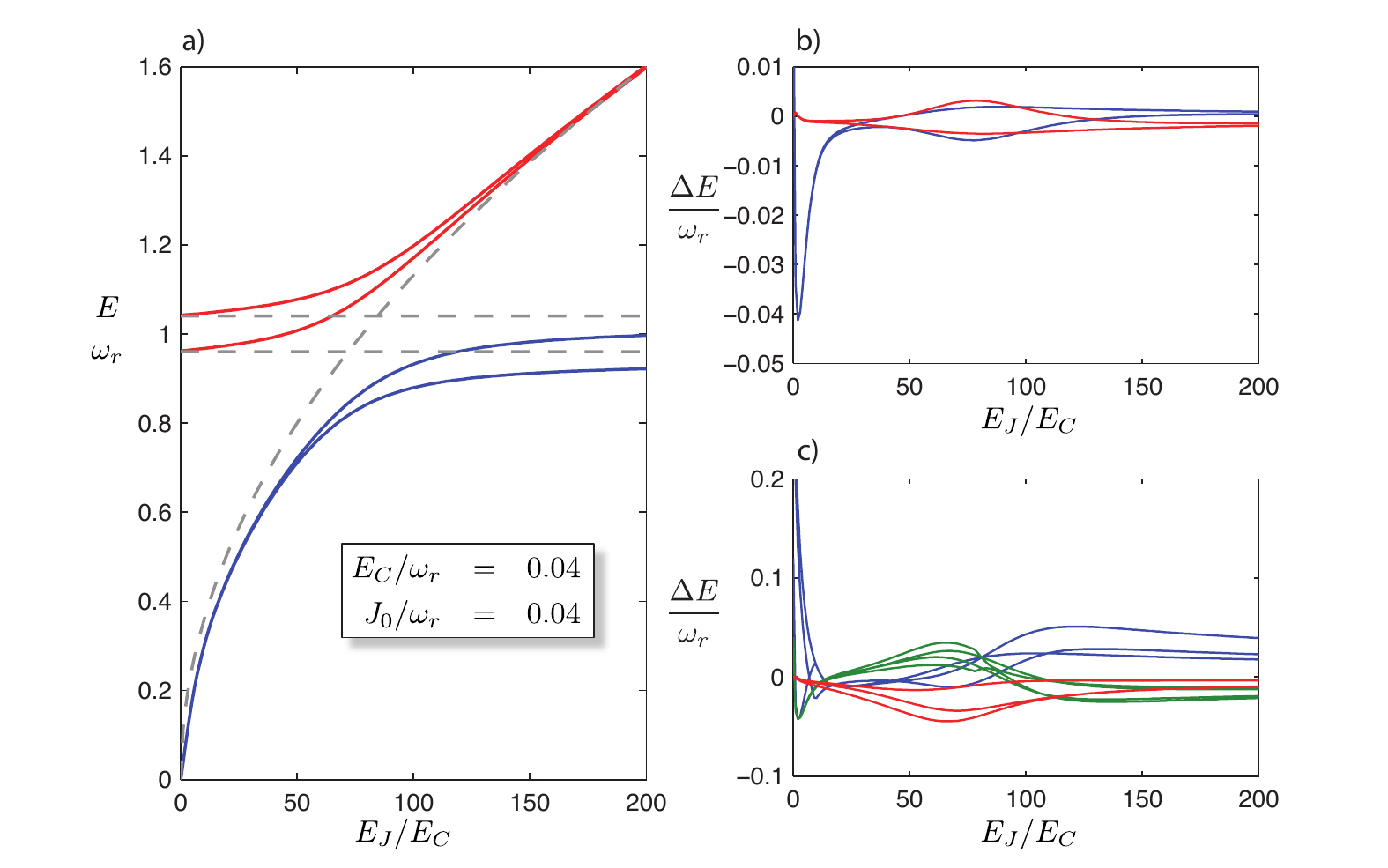}
\caption{The accuracy of our approximations. Plotted are the first four excited eigenvalues of the full Hamiltonian in figure a). Eigenvalues approximated by the $c_-$ polaritons are plotted in blue and eigenvalues approximated by the $c_+$ polaritons in red. Figures b) and c) show the differences between the eigenenergies of the full Hamiltonian $H$ and the Bose-Hubbard Hamiltonian $H^{(3)}$, $\Delta E$, in the one excitation and two excitation subspace respectively. Differences involving eigenstates containing $c_-$ polaritons are plotted in blue and differences involving eigenstates containing $c_+$ polaritons are plotted in red. Differences of eigenvalues of the full Hamiltonian and eigenvalues of the Bose-Hubbard Hamiltonian with mixed $c_+$ and $c_-$ parts are plotted in green}
\label{fig:FinalApproxJ}
\end{figure}

Both Hamiltonians conserve the total number of excitations and we can diagonalize them in each subspace with a fixed number of excitations independently.
Eigenvalues of the full Hamiltonian in the one excitation subspace are plotted in solid lines in \fref{fig:FinalApproxJ}a. Without qubits, the Hamiltonian of the two resonators has eigenmodes $a_{\pm} = (a_{1} \pm a_{2} ) / \sqrt{2}$. In  \fref{fig:FinalApproxJ} a) we also plotted the energies of these eigenmodes of the two coupled empty stripline resonators marked by two horizontal dash-dotted gray lines and the eigenenergy of the transmon qubit marked by a dash-dotted gray line. 

The two eigenenergies approximated  by the $c_-$-polariton mode (blue lines in \fref{fig:FinalApproxJ} a) evolve from the transmon qubit energy for small values of $E_J/E_C$ to the energies of the two stripline resonator states for large values of $E_J/E_C$ thereby confirming our earlier comment that the $c_-$-polaritons evolve from pure qubit excitations to photonic excitations. The eigenenergies of the $c_-$ polaritons are degenerate for small values of $E_J/E_C$ because the transmon qubits decouple from the stripline resonators and thereby also from each other.
In \fref{fig:FinalApproxJ} b) we plot differences between the eigenenergies of the full Hamiltonian and the respective eigenenergies of the Bose-Hubbard Hamiltonian. For small values of $E_J/E_C$ we find aberrations due to the error we make in approximating the Hamiltonian of the transmon qubit $H_{1-site,i}\to H^{(1)}_{1-site,i}$. There are also aberrations in the anticrossing area which are due to the neglected interactions between the $c_+$- and $c_-$-polaritons. 
 
For the two eigenenergies of the $c_+$-polaritons (red lines in \fref{fig:FinalApproxJ} a) there is a rather similar scenario. They approximate the two eigenenergies of the full Hamiltonian that are purely photonic for small $E_J/E_C$ and evolve into qubit excitations as the ratio of $E_J/E_C$ increases. There are aberrations in the anticrossing area between the eigenenergies of the full Hamiltonian and the Bose-Hubbard Hamiltonian for the $c_+$-polaritons because of the neglected interactions between the $c_-$- and $c_+$-polaritons, plotted in \fref{fig:FinalApproxJ} b) but there is no aberration for small values of $E_J/E_C$ caused by errors made in the transmon qubit Hamiltonian because the $c_+$-polaritons are pure photonic excitations for small values of $E_J/E_C$. 

Additionally to the differences between the eigenenergies of the full Hamiltonian and the Bose-Hubbard Hamiltonian in the one excitation subspace we plotted the differences in the two excitation subspace in \fref{fig:FinalApproxJ} c). These eigenenergies can be grouped for the Bose-Hubbard Hamiltonian according to the distribution of excitations among the two polariton species. Differences of eigenenergies for states with two $c_-$ polaritons are plotted in blue, for two $c_+$ polaritons in red and for one $c_-$ polariton and one $c_+$ polariton in green. In the two excitation subspace we have similar findings as in the single excitation subspace. There are aberrations for the anti-crossing area because of the neglected intermode polariton exchange interaction. In addition, states containing $c_-$ polaritons have aberrations for small values of $E_J/E_C$ due to the approximations of the transmon Hamiltonian whereas $c_+$ polaritons do not.
  
Therefore the Bose-Hubbard Hamiltonian for the $c_+$ polaritons mimics the behaviour of the full Hamiltonian for the full range of $E_J/E_C$, provided the intersite coupling $J_0$ is at most of the order of the on-site nonlinearity $E_C$ and the polariton densities are not to high. To conclude: In a driven dissipative setup where we selectively excite the $c_+$-Polaritons we do have a quantum simulator for a Bose-Hubbard Hamiltonian. 

\section{Polariton statistics in the driven dissipative regime}\label{sec:g2}

\begin{figure}
\includegraphics{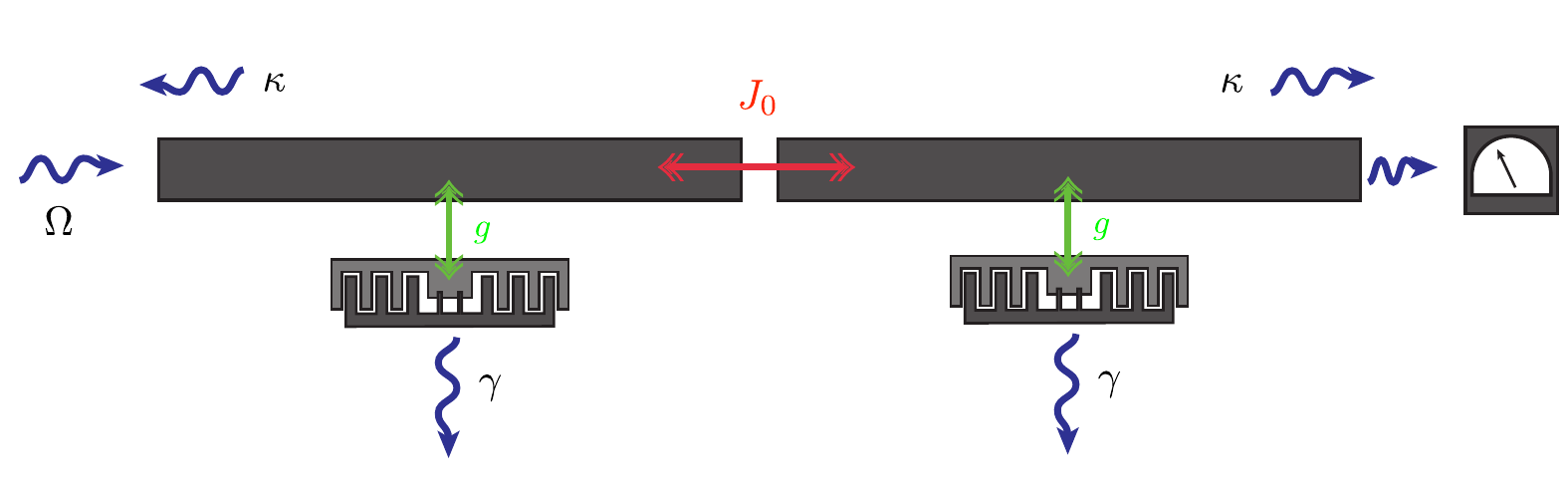}
\caption{Sketch of experimental setup consisting of two stripline resonators coupled to transmon qubits with spectroscopic measurement technique: the first stripline resonator is excited by coherent microwave source and the output field at the second cavity is monitored. Stripline resonators  decay at rate $\kappa$ mainly because of the finite reflectivity of the coupling capacitances and the transmon qubits decay Purcell enhanced at rate $\gamma$ into modes not confined by the cavity.}
\label{fig:ExperimentalSetup}
\end{figure}

In this section we make use of the above explained mapping of the full Hamiltonian $H$ to a two component Bose-Hubbard Hamiltonian $H^{(3)}$ and consider a chain of coupled resonators, where we coherently drive the first resonator and adjust the microwave drive frequency to selectively excite the $c_+$-polaritons. In the driven dissipative regime we expect to explore new physics that go beyond the equilibrium features that are commonly examined in many body physics. We thus calculate the polariton density and the density-density correlations $g^{(2)}$ in a master equation approach and analyse the dependencies on the system parameters $J_+$,$U_+$ and the Rabi frequency of the microwave drive $\Omega$.

First experimental realisations of coupled stripline resonators are expected to consist of only a few resonators. To closely approximate the expected experiments and to speed up numerical calculations, we thus focus on a minimal chain of only two resonators. More specifically, we consider two stripline resonators coupled to transmon qubits, where the first stripline resonator is driven by a microwave source and the output signal of the second cavity is monitored as a function of the microwave drive frequency and the ratio of $E_J/E_C$ which can be controlled by applying an external magnetic flux to the transmon qubits c.f. \fref{fig:ExperimentalSetup}. This setup and very similar setups are currently investigated in experiments for example \cite{2010arXiv1003.2734J}, and the spectroscopic measurement technique proposed here has already been demonstrated in single site experiments for example in \cite{Wallraff:2004rz}. 

The ouput fields are linear functions of the intra-cavity field in the second resonator and thus show the same particle statistics.
We therefore calculated the polariton density and the $g^{(2)}$-function for the second cavity. To do this, we use a master equation approach in which each element, the stripline resonators and the transmon qubits, couple to separate environments with decay rates denoted $\kappa$ for the stripline resonators and $\gamma$ for the transmon qubits.   Absolute values can for example be extracted from \cite{Bishop:2009kx} where $\gamma=3.7\mathrm{MHz}$ . Decay of the stripline resonator is due to the finite transparency of the coupling capacitors at both ends of the resonators and decay rates for example in \cite{Fink:2008sf} are $\kappa=5.7\mathrm{MHz}$. Both environments, for the transmon qubit and the stripline resonator, are assumed to be in a vaccum state which is a valid assumption at typical temperatures for circuit QED experiments of $T=15 mK$. Therefore in a master equation for a Hamiltonian expressed in the operators for the resonator field mode $a$ and the transmon qubit $b$ the dissipators read,
\begin{equation*}
\frac{\kappa}{2}\mathcal{D}[a,a]\rho+\frac{\gamma}{2}\mathcal{D}[b,b]\rho\,,
\end{equation*}
with
\begin{equation*}
\mathcal{D}[A,B]\rho=2A\rho B^{\dagger}-\left(A^{\dagger}B\rho+\rho A^{\dagger}B\right)\,.
\end{equation*}
These can be cast into dissipators expressed in the polariton modes $c_+$ and $c_-$,
\begin{eqnarray}
\frac{\kappa}{2}\mathcal{D}[a,a]\rho+\frac{\gamma}{2}\mathcal{D}[b,b]\rho&=&\Gamma_{c_+}\mathcal{D}[c_+,c_+]\rho+\Gamma_{c_-}\mathcal{D}[c_-,c_-]\rho+\nonumber\\
&&+\Lambda\left(\mathcal{D}[c_+,c_-]+\mathcal{D}[c_-,c_+]\right)\rho\,,\nonumber
\end{eqnarray}
with
\begin{eqnarray}
\Gamma_{c_+}&=&\frac{\kappa\cos^2(\theta)+\gamma\sin^2(\theta)}{2}\nonumber\\
\Gamma_{c_-}&=&\frac{\kappa\sin^2(\theta)+\gamma\cos^2(\theta)}{2}\nonumber\\
\Lambda&=&\sin(\theta)\cos(\theta)\frac{\kappa-\gamma}{2}\,.\nonumber
\end{eqnarray}
In the driven dissipative case where we selectively excite the polariton $c_+$ mode we can neglect the dissipators of the $c_-$ polaritons $\mathcal{D}[c_-,c_-]$ and  the mixed dissipators $\mathcal{D}[c_+,c_-]$ and $\mathcal{D}[c_-,c_+]$. 
With these assumptions the master equation for a two site chain of the polariton $c_+$-mode reads,
\begin{eqnarray}
\frac{d\rho}{dt}&=&\mathrm{i}\left[\rho,\tilde{H}_{c_+}\right]+\Gamma_{c_+}\left(\mathcal{D}[c_{+,1},c_{+,1}]\rho+\mathcal{D}[c_{+,2},c_{+,2}]\rho\right)\label{eq:MasterEquation}\\
\tilde{H}_{c_+}&=&\Omega\cos(\theta)\left(c_{+,1}^{\dagger}+c_{+,1}\right)+H_{c_+}\,.\nonumber
\end{eqnarray} 
To solve this master equation numerically we use it to derive the coupled equations of motion for the expectation values of normally ordered moments of the creation and annihilation operators $c_{+,1}^{\dagger}$, $c_{+,1}$, $c_{+,2}^{\dagger}$ and $c_{+,2}$,
\begin{equation*}
\frac{d}{dt}\left\langle c_{+,1}^{\dagger n}c_{+,2}^{\dagger m}c_{+,1}^k c_{+,2}^l\right\rangle=tr\left[\dot{\rho}c_{+,1}^{\dagger n}c_{+,2}^{\dagger m}c_{+,1}^k c_{+,2}^l\right]
\end{equation*} 
We truncate this set of coupled equations by omitting couplings to mean values with $n+m+k+l$ bigger than some $n_{\mathrm{max}}$ and solve the reduced set of equations of motion. To confirm the accuracy of our approach, we test its convergence with increasing $n_{\mathrm{max}}$. That is, we repeat the procedure for $n_{\mathrm{max}}\to n_{\mathrm{max}}+1$, compare the results and increase the value for $n_{\mathrm{max}}$ in case both results differ by more than some required threshold value. The advantage with respect to a method that truncates the Hilbert space at some maximal number of excitations,  is that our method becomes exact in the limit where the Hamiltonian becomes harmonic which is the case for small values of $E_J/E_C$. Moreover we experience a substantial decrease in cpu-time for this method. 
\subsection{Polariton density}

\begin{figure}
\includegraphics{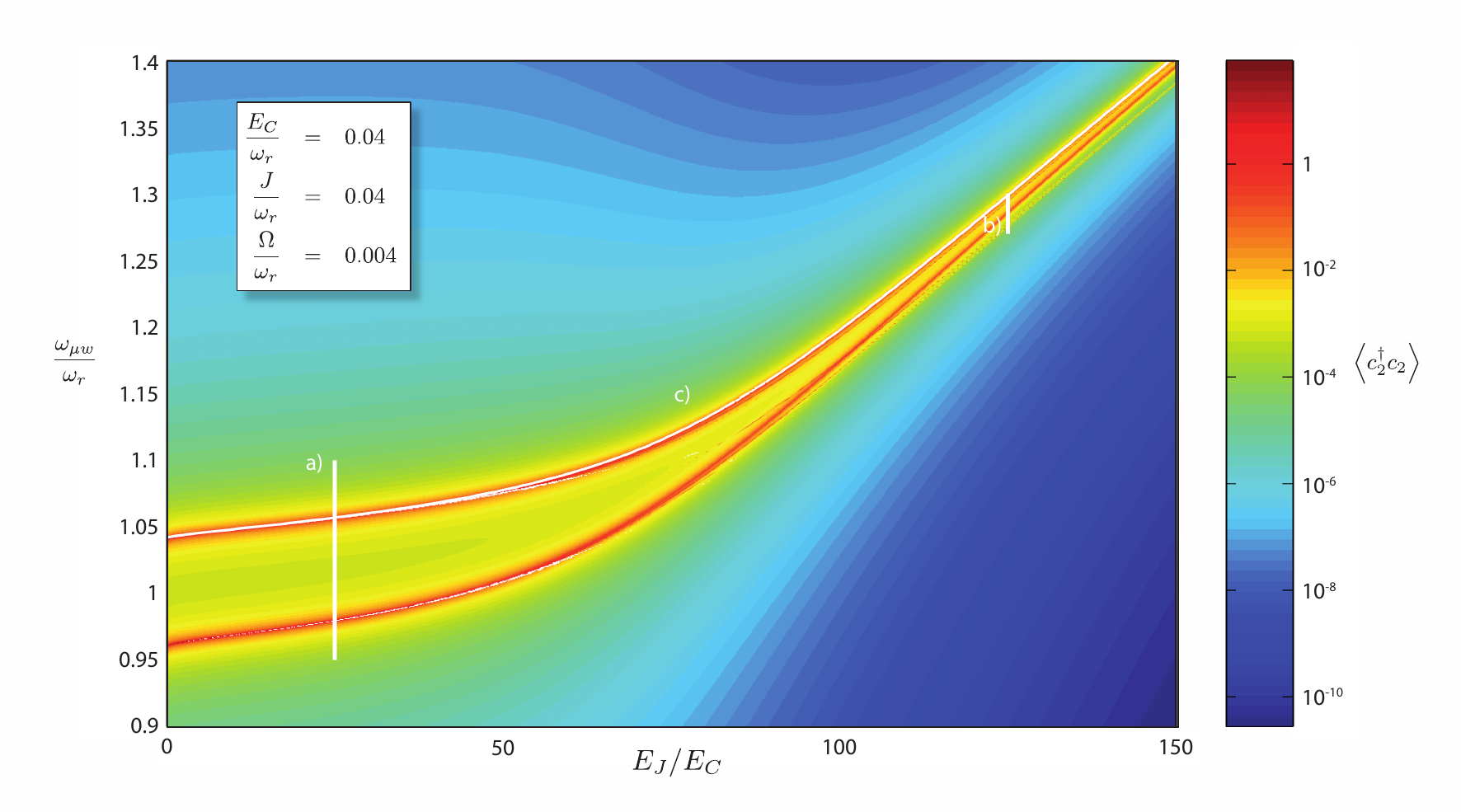}
\caption{Logarithmic density plot of the polariton density $\left\langle c_2^{\dag}c_2\right\rangle$ in the last cavity plotted against $E_J/E_C$ and the frequency of the microwave drive in units of the frequency of the stripline resonator $\omega_{\mu w}/\omega_r$. Resonances in the density of polaritons arise where the microwave frequency matches one of the transition frequencies of the non-driven conservative system Hamiltonian $H_{c_+}$}
\label{fig:densityDensity}
\end{figure}

We are interested in the field particle statistics in the driven dissipative regime and its dependencies of the on-site nonlinearity $U_+$, the intersite coupling $J_+$ and the strength of the microwave drive $\Omega$. We therefore first consider the density of $c_-$ polaritons in the last resonator.
\Fref{fig:densityDensity} shows the density of $c_+$ polaritons, $\left\langle c_2^{\dag}c_2\right\rangle$, in the second cavity as a function of the ratio $E_J/E_C$ and the microwave drive frequency $\omega_{\mu w}$. The density of polaritons in the last cavity exhibits resonances when the microwave drive frequency matches one of the transition energies of the undriven conservative system Hamiltonian $H_{c_+}$ and decreases rapidly because of the small decay rate $\Gamma$. One can clearly see the resonances due to transitions driven between the groundstate and eigenenergies in the one excitation subspace plotted in \fref{fig:FinalApproxJ} a). Transitions from the groundstate into a two excitation state are much weaker owing to the finite Rabi frequency of the microwave drive $\Omega$. 

\subsection{Density-density correlations}

\begin{figure}
\includegraphics{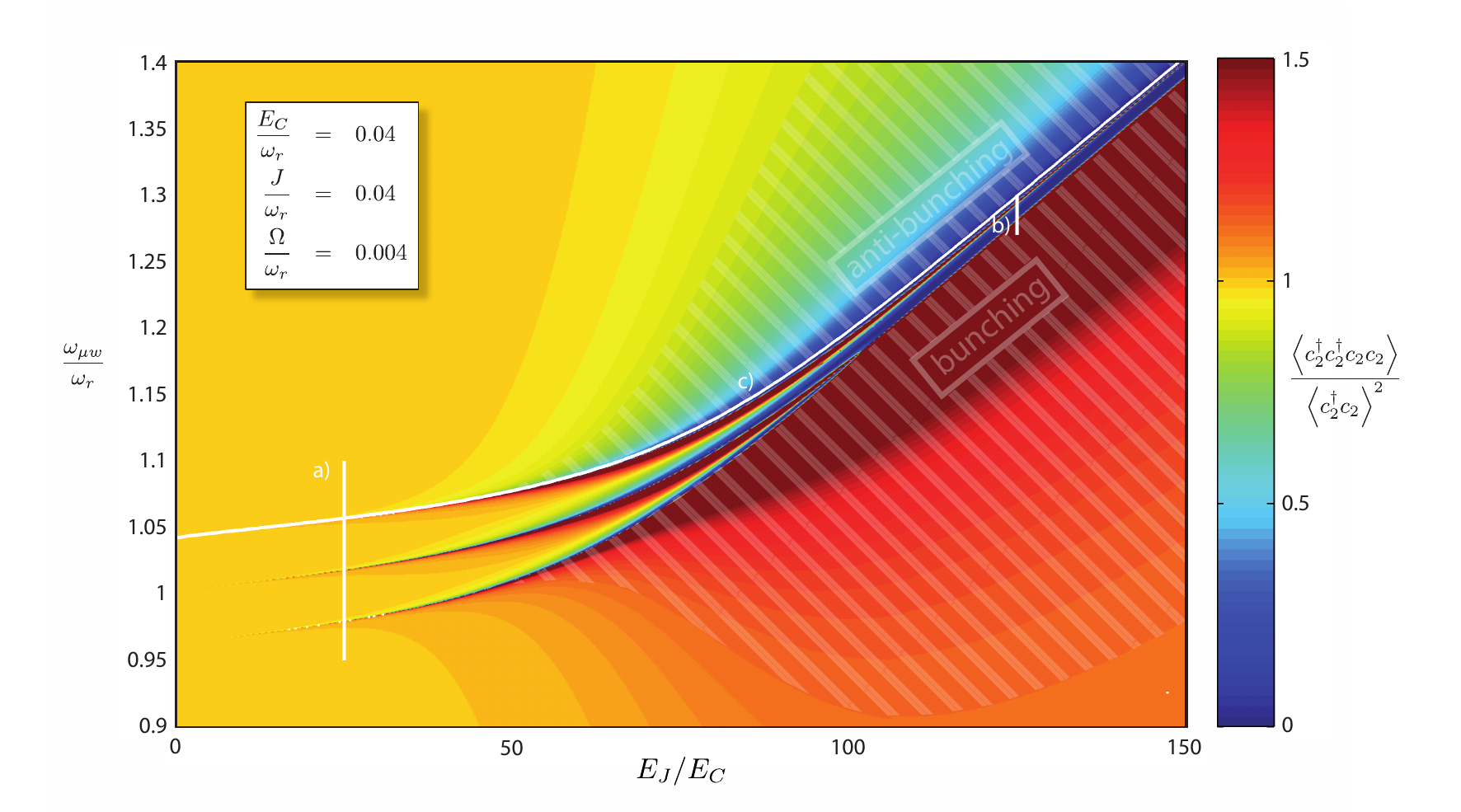}
\caption{Density plot of the $g^{(2)}$-function of the polariton mode in the second cavity plotted against  $E_J/E_C$ and the frequency of the microwave drive in units of the frequency of the stripline resonator $\omega_{\mu w}/\omega_r$. For increasing nonlinearity the $g^{(2)}$-function shows bunching regions for red-detuned microwave drive with respect to the energy of  1-excitation states and anti-bunching regions for blue-detuned microwave drive.}
\label{fig:g2density}
\end{figure}

We now consider the density-density correlations $g^{(2)}$ in the last resonator. The $g^{(2)}$-function is a quantity that describes the likelihood to measure two photons at the same place. The $g^{(2)}$ function of the last resonator is the normalized meanvalue of the second order moment of the field operators in the last resonator, 
\begin{equation*}
g^{(2)}(c_{+,2})=\frac{\left\langle c_{+,2}^{\dag}c_{+,2}^{\dag}c_{+,2}c_{+,2}\right\rangle}{\left\langle c_{+,2}^{\dag}c_{+,2}\right\rangle^2}\,.
\end{equation*}
Classical, thermal fields have $g^{(2)}$-values larger or equal to unity with the coherent field exhibiting a $g^{(2)}$-value of 1. A $g^{(2)}$-value below 1, meaning that the photons are anti-bunched, is a sufficient condition to call the field quantum mechanical in the sense that there is no classical field showing the same results in measurements of the $g^{(2)}$-function. With recently developed refinements of microwave measurement techniques \cite{2010arXiv1001.3669M,Bergeal:2010fk}, measurements of $g^{(2)}$-functions in circuit QED are now becoming feasible.
 
\begin{figure}
\includegraphics{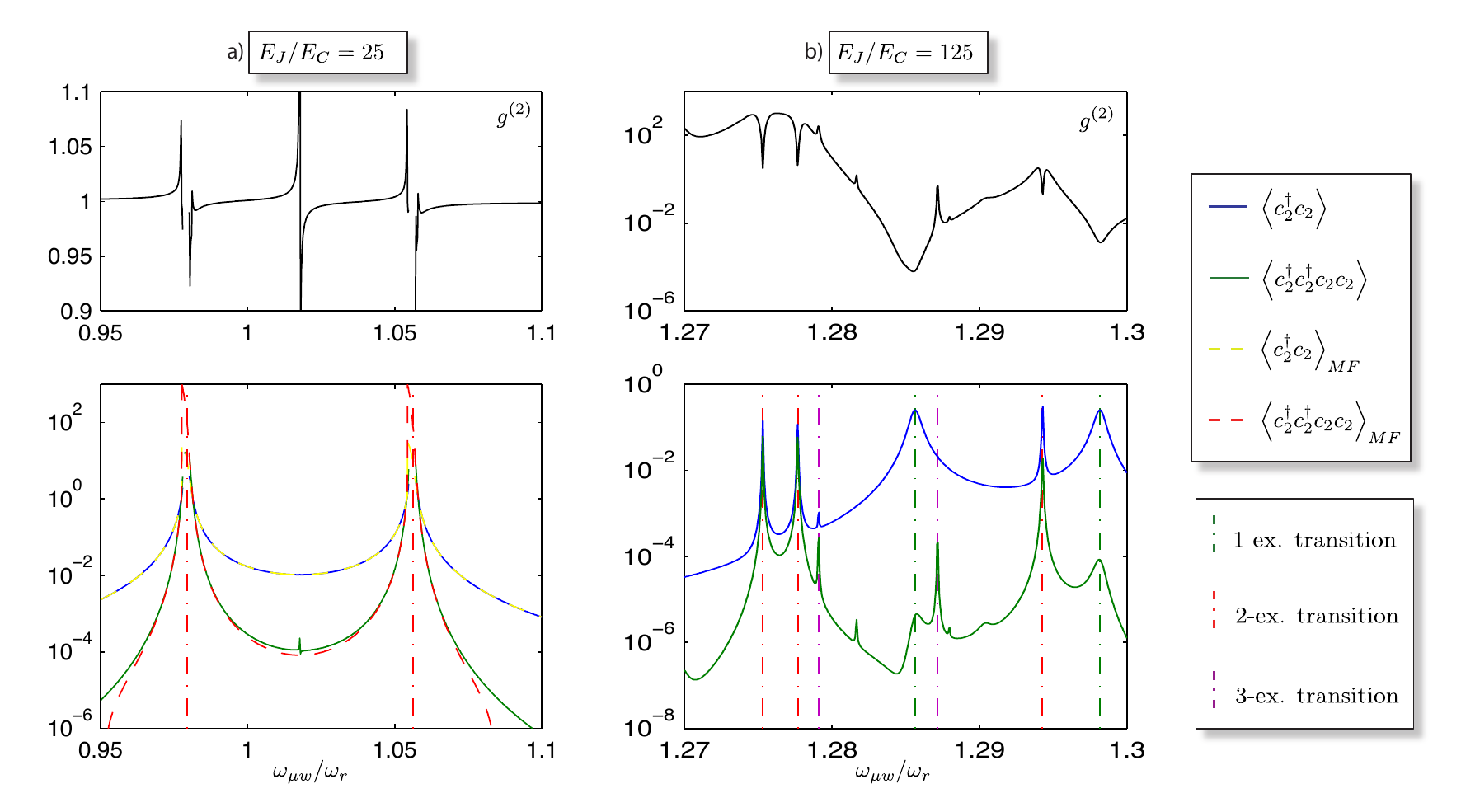}
\caption{Plots of the $g^{(2)}$-function, the density of polaritons $\left\langle c_2^{\dagger}c_2\right\rangle$ and the second order moment $\left\langle c_2^{\dagger} c_2^{\dagger}c_2c_2\right\rangle$ in the second cavity for special values of $E_J/E_C$ are shown. For all plots we have chosen the intersite coupling constant and the on-site nonlinearity to be  $J_0/\omega_r=0.04$ and $E_C/\omega_r=0.04$ and the decay rates of transmon qubit and stripline resonator to be $\gamma/\omega_r=0.00008$ and $\kappa/\omega_r=0.00004$ respectively but we applied different Rabi frequencies of the microwave drive: for a) $\Omega/\omega_r=0.004$, and for b) $\Omega/\omega_r=0.001$. Plotted are results obtained by numerical calculation of the masterequation in solid lines and results obtained by a mean field approach with an exact single-site solution in dashed lines. Eigenenergies of the system without dissipation and driving are signalized by vertical dash-dotted lines. For $E_J/E_C=25$ one can see clearly separated resonances for the symmetric and antisymmetric states $d_{\pm}^{\dagger}\left|00\right\rangle$ and a two photon resonance for the state $d_{+}^{\dagger} d_{-}^{\dagger} \left|00\right\rangle$. The shape of the resonances at $d_{+}^{\dagger}\left|00\right\rangle$ and $d_{-}^{\dagger}\left|00\right\rangle$ are reproduced by the meanfield approximation and therefor a single mode feature. The two photon resonance for the sate $d_{+}^{\dagger} d_{-}^{\dagger} \left|00\right\rangle$ is not reproduced by the meanfield approach since it does not correctly incorporate the interactions between $d_{+}^{\dagger}$ and $d_{-}^{\dagger}$ modes. For $E_J/E_C=125$ multiple resonances determined by the eigenenergies of the system without dissipation and driving arise .}
\label{fig:tracePlots}
\end{figure}

\begin{figure}
\includegraphics{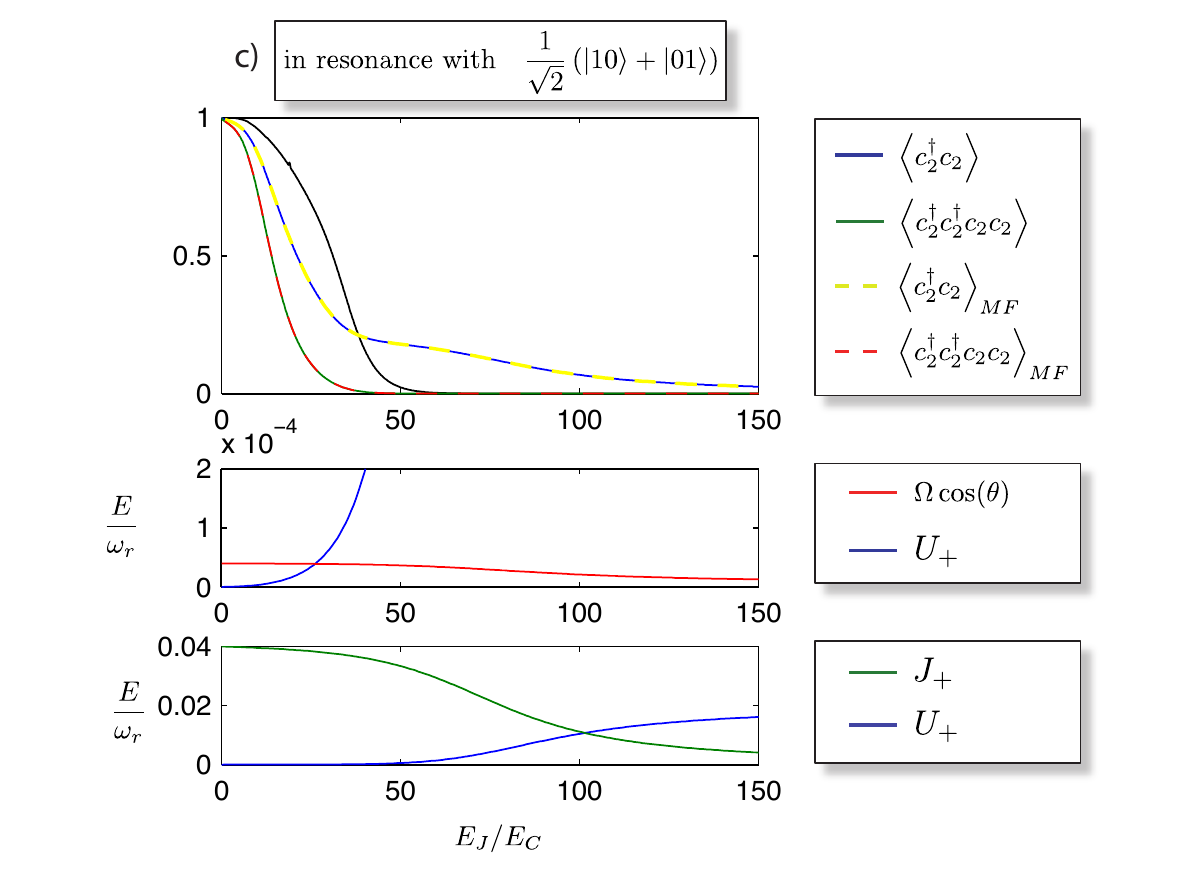}
\caption{A plot of the $g^{(2)}$-function, the density of polaritons $\left\langle c_2^{\dagger}c_2\right\rangle$ and the second order moment $\left\langle c_2^{\dagger} c_2^{\dagger}c_2c_2\right\rangle$ in the second cavity is shown. The microwave drive frequency is chosen to match the transition from the ground state to the symmetric 1-excitation eigenstate. For all plots we have chosen the intersite coupling constant and the on-site nonlinearity to be  $J_0/\omega_r=0.04$ and $E_C/\omega_r=0.04$, the decay rates for transmon qubit and stripline resonator to be $\gamma/\omega_r=0.00004$ and $\kappa/\omega_r=0.00008$ respectively and the Rabi frequency of the microwave drive to be $\Omega/\omega_r=\Gamma/\omega_r=0.00004$. Plotted are results obtained by numerical calculation of the masterequation in solid lines and results obtained by a mean field approach with an exact single-site solution in dashed lines. In resonance to the symmetric state $g^{(2)}$ shows a transition from uncorrelated coherent field particle statistics to anti-bunched correlated field particle statistics. The transition from coherent to anti-bunched is determined by the interplay of the Rabi frequency of the microwave drive and the on-site nonlinearity.}
\label{fig:ResonancePlot}
\end{figure}

In \fref{fig:g2density} we plotted the $g^{(2)}$-function of the field in the last stripline resonator. To get a more detailed insight of the processes leading to a $g^{(2)}$-value for specific parameters we plotted the $g^{(2)}$-function along special values of the microwave drive frequency and the ratio $E_J/E_C$ marked by white lines in \fref{fig:g2density}. Figures \ref{fig:tracePlots} and \ref{fig:ResonancePlot} show the results for the different paths, denoted by a), b), and c) in the density plot of the $g^{(2)}$-function in \fref{fig:g2density}, for the $g^{(2)}$-function as well as the corresponding values for the density of polaritons in the last cavity, $\langle c_{+,2}^{\dagger}c_{+,2}\rangle$, and the second order moment, $\langle c_{+,2}^{\dagger}c_{+,2}^{\dagger}c_{+,2}c_{+,2}\rangle$.

For small values of $E_J/E_C$ our system is basically linear because the nonlinearity $U_{+}/2 = (E_C/2)\sin^4(\theta)$ is negligible. A harmonic field mode driven by a coherent source is in a coherent state. Therefore the $g^{(2)}$-function is equal to one for small values of $E_J/E_C$. As the nonlinearity grows for increasing values of $E_J/E_C$ the $g^{(2)}$-function plotted against the ratio of $E_J/E_C$ and the frequency of the microwave drive $\omega_{\mu w}$ gets more structured.
In the density plot of the $g^{(2)}$-function in \fref{fig:g2density} we can identify resonances where the frequency of the microwave drive matches the eigenenergies of the unperturbed system without microwave drive and dissipation. These resonances manifest themselves as  separating lines between bunching regions (values of $g^{(2)}>1$) and anti-bunching regions (values of $g^{(2)}<1$).

To understand the origin of these separating lines, it is illustrating to analyze our system in terms of a symmetric mode, $d_+$, and an antisymmetric mode, $d_-$, where
\begin{equation*}
d_{\pm}=\frac{1}{\sqrt{2}}\left(c_{+,1}\pm c_{+,2}\right)
\end{equation*}
rather than the two localized modes $c_{+,1}$ and $c_{+,2}$. In terms of $d_+$ and $d_-$ the Hamiltonian reads,
\begin{eqnarray}
H_{c_+}&=&\left(\omega_+-J_+\right)d_+^{\dagger}d_++\left(\omega_++J_+\right)d_-^{\dagger}d_--\frac{U_+}{4}\left(d_+^{\dagger}d_+^{\dagger}d_+d_++d_-^{\dagger}d_-^{\dagger}d_-d_-\right)\nonumber\\
&&-U_+ d_+^{\dagger}d_+d_-^{\dagger}d_--\frac{U_+}{4}\left(d_+^{\dagger}d_+^{\dagger}d_-d_-+h.c.\right)\,.\label{eq:CmHamiltonian}
\end{eqnarray}
The Hilbert space of the Hamiltonian $H_{c+}$ can be described by two different bases, states that are labeled by the number of excitations in the collective modes,
\begin{equation*}
\frac{(d_+^{\dagger})^n}{\sqrt{n!}}\frac{(d_-^{\dagger})^m}{\sqrt{m!}}\left|00\right\rangle=\left|nm\right\rangle_{cm}\,,
\end{equation*}
or states that are labeled by the number of excitations in the localized modes,
\begin{equation*}
\frac{(c_{+,1}^{\dagger})^n}{\sqrt{n!}}\frac{(c_{+,2}^{\dagger})^m}{\sqrt{m!}}\left|00\right\rangle=\left|nm\right\rangle_{s}\,.
\end{equation*}
 The lines separating bunching and anti-bunching regions in figure \ref{fig:g2density} can now be identified with the energies of the 1 excitation states,
 \numparts
\begin{eqnarray}\label{eq:1exstate}
d_+\left|00\right\rangle&=&\left|10\right\rangle_{cm}=\frac{1}{\sqrt{2}}\left(\left|10\right\rangle_{s}+\left|10\right\rangle_{s}\right)\\
d_-\left|00\right\rangle&=&\left|10\right\rangle_{cm}=\frac{1}{\sqrt{2}}\left(\left|10\right\rangle_{s}-\left|10\right\rangle_{s}\right)
\end{eqnarray}
 and the energy of a 2-excitation state,
 \begin{equation}
d_+^{\dagger}d_-^{\dagger}\left|00\right\rangle=\left|11\right\rangle_{cm}=\frac{1}{\sqrt{2}}\left(\left|20\right\rangle_s-\left|02\right\rangle_s\right)\,.
\end{equation} 
\endnumparts
\begin{figure}
\includegraphics{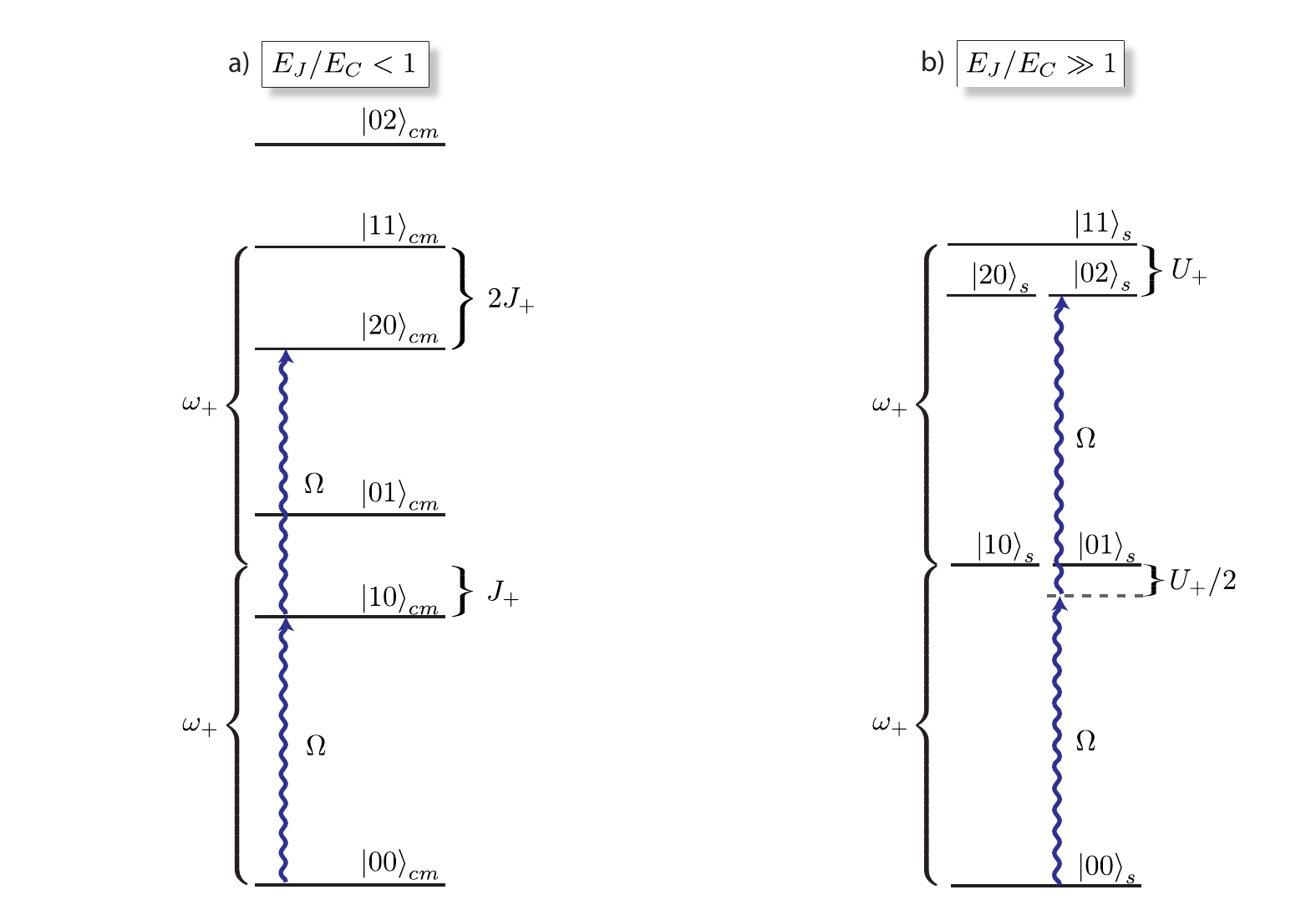}
\caption{Sketch of the energy spectrum of the Bose-Hubbard Hamiltonian $H_{c_+}$ for a two site model for vanishing nonlinearity $U_+$ compare a) and vanishing intersite coupling $J_+$ compare b). For vanishing nonlinearity a microwave drive can drive multiple transitions leading to a coherent state. Contrary to the linear case for strong nonlinearity one can only drive a transition between two distinct states as the energy differences between the eigenenergies aren't degenerate any more}
\label{fig:2siteSpectrum}
\end{figure}
To understand the origin of the anti-bunching regions for a microwave drive that is blue detuned with respect to the energies of the states (\ref{eq:1exstate}{\it-c}) and the bunching regions for a red detuned microwave drive one has to consider the spectrum of the Hamiltonian $H_{c_+}$. 

For small nonlinearity, that is for values of $E_J/E_C < 50$, the Hamiltonian \eref{eq:CmHamiltonian} reduces to a Hamiltonian for two uncoupled harmonic oscillators described by the modes $d_+$ and $d_-$ with energies $\omega_+-J_+$ and $\omega_++J_+$ respectively. The eigenenergies in this situation are shown in \fref{fig:2siteSpectrum} a). A microwave drive with frequency $\omega_+-J_+$ as depicted in \fref{fig:2siteSpectrum} not only drives the transition from the groundstate to the first excited state of the symmetric collective mode $\left|0\,0\right\rangle_{cm}\to\left|1\,0\right\rangle_{cm}$ but also all other transitions to higher excited states $\left|n\,0\right\rangle_{cm}\to\left|n+1\,0\right\rangle_{cm}$. As a result the steady state in this situation is always the coherent state exhibiting a $g^{(2)}$-value of $1$. For slightly increased values of the nonlinearity that remain in the range $U_+<\Gamma_{c_+}$, the system can still be described in terms of two weakly interacting collective modes. But the symmetric as well as the antisymmetric mode are subject to the nonlinearity and an intermode interaction, c.f. Hamiltonian \eref{eq:CmHamiltonian}. This can be seen in \fref{fig:tracePlots} a) where we plotted the $g^{(2)}$-values that deviate from the value of a coherent field. The $g^{(2)}$-function shows anti-bunching regions for blue detuned microwave drive with respect to the energies of the states (\ref{eq:1exstate}{\it-c}) and bunching regions for red detuned microwave drive.
To gain insight into the underlying physical principles in this situation we calculated the density $\langle c_{+,2}^{\dagger}c_{+,2}\rangle$ and the second symmetric moment $\langle c_{+,2}^{\dagger}c_{+,2}^{\dagger}c_{+,2}c_{+,2}\rangle$ by an iterative meanfield approach to solve the master equation \eref{eq:MasterEquation} with the Hamiltonian written as in \eref{eq:CmHamiltonian}. Operator mean values of a single driven dissipative mode with Kerr nonlinearity can be computed exactly  \cite{0305-4470-13-2-034} and we expand this model in a meanfield way to incorporate the denisty-density coupling. With this method we get good agreement with the numerical exact values for the density in the last cavity and are able to compute values for the polariton density close to the systems eigenenergies (\ref{eq:1exstate}{\it-c}) where our numerical approach fails to converge. For details about the method please see \ref{app:MeanField}. These results support our assertion that the system can be described by weakly interacting collective modes in the limit of small nonlinearities $U_+$. In \fref{fig:tracePlots} a) numerically exact values are plotted in solid lines and values obtained by the above mentioned mean field method are plotted in dashed lines.

For strong nonlinearity $U_+$ and small intersite coupling $J_+$, that is for values of $E_J/E_C > 50$, the $c_{+}$-polaritons become transmon excitations and the Hamiltonian $H_{c_+}$ splits into two parts describing the first and the second transmon qubit respectively. Here the collective modes $d_{+}$ and $d_{-}$ no longer decouple and the localized modes $c_{1}$ and $c_{2}$ become a more appropriate description of the system. The eigenenergy spectrum in this situation is shown in \fref{fig:2siteSpectrum} b). The main difference to the spectrum without nonlinearity is that the microwave drive can't be adjusted to drive multiple transitions. In order to drive the transition to the state $\left|02\right\rangle_s$ for example one has to adjust the microwave frequency to match half of the energy difference between the groundstate and the 2-excitation state  $\left|02\right\rangle_s$ because it is a two photon transition. Due to the anharmonicity of the eigenenergy spectrum no other transition can be driven. The difference of microwave frequencies needed to drive the transition from groundstate to $\left|01\right\rangle_s$ respectively $\left|02\right\rangle_s$ amounts to $U_+/2$ which is bigger than the linewidth $\Gamma_{c_+}$. To get an estimate for the value of $g^{(2)}$ we simplify our model assuming that the frequency of the microwave drive is adjusted such that it resonantly drives a transition between the groundstate of our model $\left|00\right\rangle_s$ and some excited state $\left|0n\right\rangle_s$. Provided the Rabi frequency $\Omega$ and the loss rates $\kappa$ and $\gamma$ are all small compared to the frequency separation between different resonance lines, the system can then be modeled by a two level system consisting of the groundstate of our model $\left|00\right\rangle_s$ and the excited state $\left|0n\right\rangle_s$. In this situation the maximal occupation inversion one could get in the steady state is, 
\begin{equation*}
\rho_{max}=\frac{1}{2}\left(\left|00\right\rangle\left\langle 00\right|+\left|0n\right\rangle\left\langle 0n\right|\right)\,,
\end{equation*}
and the $g^{(2)}$-value for this density matrix would be
\begin{equation*}
g^{(2)}=\frac{\text{tr}\left[\rho_{max}c_{+,2}^{\dagger}c_{+,2}^{\dagger}c_{+,2}c_{+,2}\right]}{\left(\text{tr}\left[\rho_{max}c_{+,2}^{\dagger}c_{+,2}\right]\right)^2}=\frac{2n(n-1)}{n^2}
\end{equation*}
which is below one for a 1-excitation state and above one for every state containing more than 2 excitations. Therefore bunching areas arise if states containing more than 2-exciations are excited and anti-bunching areas arise if only 1-exitation states can be excited and the photons pass the setup ``one by one''. In our Bose-Hubbard model the on-site nonlinearity is negative and hence all transition frequencies to states containing more than two excitations are red detuned with respect to transition frequencies to states containing only one excitation (\ref{eq:1exstate}{\it-c}). This is why bunching areas arise for red detuned microwave drive and anti bunching areas arise for blue detuned microwave drive.

If we adjust the microwave drive frequency for every value of $E_J/E_C$ to match the eigenfrequency of the antisymmetric 1-excitation state we get the transition from a perfectly uncorrelated field with $g^{(2)}=1$ to strongly correlated, anti-bunched field statistics with $g^{(2)}<1$ see \fref{fig:ResonancePlot}. For a quantum phase transition of the ground state of the Bose-Hubbard Hamiltonian one would expect this transition as a consequence of the interplay of the intersite hopping $J_+$ and the on-site nonlinearity $U_+$. For the driven dissipative system we observe that the interplay between the Rabi frequency of the microwave drive $\Omega\cos(\theta)$ and the on-site nonlinearity $U_+$ determines the particle statistics. This can be seen in \fref{fig:ResonancePlot} where we plotted the $g^{(2)}$-function and the intersite coupling, on-site nonlinearity and the Rabi frequency of the microwave drive.

\section{Summary}
We have shown that a chain of capacitively coupled stripline resonators each coupled to a transmon qubit can be described by a Bose-Hubbard Hamiltonian for two species of polaritons. The validity of our approach has been checked for realistic parameters of the transmon qubits and stripline resonators and for low densities of polaritons. In a driven dissipative regime where a microwave source coherently drives the first cavity one can selectively excite only one species of polaritons and investigate the properties of a driven dissipative Bose-Hubbard model. We calculated the density and the $g^{(2)}$-function of the polaritons in the last resonator of a two site setup and investigated their dependencies on the microwave drive, the intersite coupling and the on-site nonlinearity. For vanishing nonlinearity the $g^{(2)}$-function is approximately equal to unity indicating a coherent field. With increasing nonlinearity bunching and anti-bunching areas arise depending on the frequency of the microwave drive. For a microwave drive that is in resonance with a transition to a state with one excitation, the polaritons are anti-bunched. If, on the other hand, the microwave drive can resonantly excite states containing more than 2 excitations in a multi-photon transition, the polaritons become bunched. If we adjust the microwave drive frequency to match one of the system's single excitation eigenenergies and compute the $g^{(2)}$-function for different values of on-site nonlinearity, intersite coupling and Rabi frequency of the microwave drive we see a transition from coherent to anti-bunched field statistics. That is, the polaritons are uncorrelated for small nonlinearity and exhibit a transition to anti-bunched behaviour as the on-site nonlinearity becomes larger than the Rabi frequency of the microwave drive. All our findings could be explored in experiments based on readily available technology.

\ack
This work is part of the Emmy Noether project 
HA 5593/1-1 funded by Deutsche Forschungsgemeinschaft (DFG).

\appendix
\section[Meanfield Approximation]{Meanfield approximation for two coupled modes with small nonlinearity}\label{app:MeanField}
We want to solve the master equation \eref{eq:MasterEquation} for small values of $E_J/E_C$. For $E_J/E_C = 25$, we find $J_{+} \gg U_{+}$ and express Hamiltonian $\tilde{H}_{c_+}$ in terms of the collective modes $d_+$ and $d_-$,
\begin{equation*}
\tilde{H}_{c_+}=\tilde{H}_{d_+}+\tilde{H}_{d_-}+H_{dd}+H_{ex}
\end{equation*}
where,
\begin{eqnarray}
\tilde{H}_{d_\pm}&=&\frac{\Omega\cos(\theta)}{\sqrt{2}}\left(d_{\pm}^{\dagger}+d_{\pm}\right)+\left(\Delta\omega_+\pm J_+\right)d_{\pm}^{\dagger}d_{\pm}-\frac{U_+}{4}d_{\pm}^{\dagger}d_{\pm}^{\dagger}d_{\pm}d_{\pm}\nonumber\\
H_{dd}&=&-U_+d_+^{\dagger}d_+d_-^{\dagger}d_-\nonumber\\
H_{ex}&=&-\frac{U_+}{4}\left(d_+^{\dagger}d_+^{\dagger}d_-d_-+h.c.\right)\,.\nonumber
\end{eqnarray}
Here $\Delta\omega_+=\omega_+-\omega_l$ is the difference between the frequency $\omega_+$ and the microwave drive frequency $\omega_l$.
Since $J_+\gg U_+$, the two collective modes are energetically far separated and we neglect $H_{ex}$ in a rotating wave approximation. In zeroth order in our approximation we neglect $H_{dd}$ as well and decouple the collective modes completely. The driven dissipative master equation for the collective modes $d_+$ and $d_-$ with Hamiltonians $H_{d_+}$ and $H_{d_-}$ can be solved exactly for steady state values, c.f. \cite{0305-4470-13-2-034}. We compute the densities of the two collective modes $\left\langle d_{\pm}^{\dagger}d_{\pm}\right\rangle=\delta_{\pm}^{(0)}$ in zeroth order approximation and use them to approximate the intermode density-density coupling,
\begin{equation*}
-U_+d_+^{\dagger}d_+d_-^{\dagger}d_-\to-\frac{U_+}{2}\left(d_+^{\dagger}d_+\delta_-^{(0)}+\delta_+^{(0)}d_-^{\dagger}d_-\right).
\end{equation*}
This way the modes are still decoupled but the density of the $d_+$-mode induces a frequency shift of the $d_-$-mode $\omega_+-J_+\to\omega_+-J_+-U_+\delta_+^{(0)}$ and vice versa. We proceed to calculate the densities of the two modes with the shifted frequencies to obtain the densities of the two modes in first order approximation $\delta_{\pm}^{(1)}$. We iterate this method until the difference between densities of consecutive order in approximation is smaller than some threshold value.
After this procedure we approximate the density in the last stripline resonator 
\begin{equation*}
\left\langle c_{+,2}^{\dagger}c_{+,2}\right\rangle\to\frac{1}{2}\left(\left\langle d_+^{\dagger}d_+\right\rangle+\left\langle d_-^{\dagger}d_-\right\rangle\right)+\text{Re}\left[\left\langle d_+^{\dagger}\right\rangle\left\langle d_-\right\rangle\right]
\end{equation*}
where we calculated the values for $\left\langle d_+^{\dagger}\right\rangle$ and $\left\langle d_-\right\rangle$ with the renormalized frequencies $\omega_++J_+-U_+\delta_-^{(n_{max})}$ and $\omega_+-J_+-U_+\delta_+^{(n_{max})}$. With this procedure we can reproduce the numerically exact values for the density in the last stripline resonator and are able to compute values for the polariton density close to the systems eigenenergies (\ref{eq:1exstate}{\it-c}) where the numerically exact method converges very slowly and becomes numerically very demanding.
\noappendix

\bibliographystyle{unsrt}

\end{document}